\documentclass{IOS-Book-Article}

\usepackage{mathptmx}
\usepackage{soul}\setuldepth{article}
\usepackage{float}
\def\hb{\hbox to 11.5 cm{}}

\usepackage{graphicx}

\usepackage{tikz}
\usepackage{tikzscale}
\usetikzlibrary{positioning, shapes.multipart, arrows.meta, backgrounds, fit, quotes}

\usepackage{orcidlink}

\definecolor{Class}{RGB}{207,165,0}      
\definecolor{ObjPro}{RGB}{0,121,186}      

\usepackage{amsmath}
\usepackage{amsfonts}
\usepackage{todonotes}
\usepackage{xspace}
\usepackage{newfloat} 
\usepackage{cleveref}
\usepackage{ntheorem}
\usepackage{stmaryrd}
\usepackage{multirow}
\usepackage{relsize}
\usepackage{booktabs}
\usepackage{ifthen}

\usepackage[edges]{forest}

\usepackage{placeins}

\theoremstyle{numberbreak}
\theorembodyfont{\normalfont}

\theoremstyle{break}
\theorembodyfont{\normalfont}

\newcommand{\nop}[1]{}

\newcommand{\secuman}{\textsc{Secuman}\xspace}
\newcommand{\riskman}{\textsc{Riskman}\xspace}

\newcommand{\ebnfeq}{\mathrel{::=}}
\newcommand{\ebnfalt}{\mathbin{\mid}}

\newcommand{\define}[1]{\textit{#1}}

\newcommand{\set}[1]{\left\{#1\right\}}
\newcommand{\tuple}[1]{\left(#1\right)}
\newcommand{\guard}{\ \middle\vert\ }
\newcommand{\card}[1]{\left\lvert#1\right\rvert}

\newcommand{\refreq}[2]{} 

\newcommand{\vdespec}{VDE Spec 90025\xspace}
\newcommand{\vdespeccite}{\vdespec~\cite{VDESpec24}\xspace}

\newcommand{\isonorm}{ISO 14971\xspace}
\newcommand{\isonormcite}{\isonorm~\cite{ISO14971}\xspace}

\tikzstyle{myborder}=[densely dashed]

\newcommand{\owl}{OWL\xspace}
\newcommand{\shacl}{SHACL\xspace}
\newcommand{\riskmanurl}[1][]{\url{https://w3id.org/riskman#1}\xspace}
\newcommand{\ontologyurl}[1][]{\riskmanurl[/ontology]}
\newcommand{\shapesurl}[1][]{\riskmanurl[/shapes]}
\newcommand{\dlfont}[1]{\ensuremath{\mathsf{#1}}}

\newcommand{\AttPro}{\dlfont{AttackerProfile}}
\newcommand{\Boolean}{\dlfont{Boolean}}
\newcommand{\ImpLev}{\dlfont{ImpactLevel}}
\newcommand{\ImpAss}{\dlfont{ImpactAssessment}}
\newcommand{\ImpAssDom}{\dlfont{ImpactAssessmentDomain}}
\newcommand{\LikFac}{\dlfont{LikelihoodFactor}}
\newcommand{\Rat}{\dlfont{Rationale}}
\newcommand{\SecRis}{\dlfont{SecurityRisk}}
\newcommand{\SecRisLev}{\dlfont{SecurityRiskLevel}}
\newcommand{\VulLev}{\dlfont{ExposureLevel}}

\newcommand{\hAttPro}{\dlfont{hasAttackerProfile}}
\newcommand{\hImpLev}{\dlfont{hasImpactLevel}}
\newcommand{\hLikFac}{\dlfont{hasLikelihoodFactor}}
\newcommand{\hRat}{\dlfont{hasRationale}}
\newcommand{\hSafImpAss}{\dlfont{hasSafetyImpactAssessment}}
\newcommand{\hSecImpAss}{\dlfont{hasSecurityImpactAssessment}}

\newcommand{\hSecRisLev}{\dlfont{hasSecurityRiskLevel}}
\newcommand{\hVulLev}{\dlfont{hasExposureLevel}}
\newcommand{\iAff}{\dlfont{isAffected}}

\newcommand{\AttTyp}{\dlfont{AttackType}}
\newcommand{\Ava}{\dlfont{availability}}
\newcommand{\Con}{\dlfont{confidentiality}}
\newcommand{\ConSpeThr}{\dlfont{ContextSpecificThreat}}
\newcommand{\DatSecAndPri}{\dlfont{DataSecurityAndPrivacy}}
\newcommand{\IdeAndAccMan}{\dlfont{IdentityAndAccessManagement}}
\newcommand{\Int}{\dlfont{integrity}}
\newcommand{\NetSec}{\dlfont{NetworkSecurity}}
\newcommand{\OpeAndMai}{\dlfont{OperationsAndMaintenance}}
\newcommand{\OpeEnv}{\dlfont{OperatingEnvironment}}
\newcommand{\PhySec}{\dlfont{PhysicalSecurity}}
\newcommand{\Pro}{\dlfont{Product}}
\newcommand{\ProGoa}{\dlfont{ProtectionGoal}}
\newcommand{\SysAndPlaSec}{\dlfont{SystemAndPlatformSecurity}}

\newcommand{\ProGoaSet}{\{\Ava,\Con,\Int\}}

\newcommand{\hAttTyp}{\dlfont{hasAttackType}}
\newcommand{\hConSpeThr}{\dlfont{hasContextSpecificThreat}}
\newcommand{\hDatSecAndPri}{\dlfont{hasDataSecurityAndPrivacy}}
\newcommand{\hIdeAndAccMan}{\dlfont{hasIdentityAndAccessManagement}}

\newcommand{\hNetSec}{\dlfont{hasNetworkSecurity}}
\newcommand{\hOpeAndMai}{\dlfont{hasOperationsAndMaintenance}}
\newcommand{\hOpeEnv}{\dlfont{hasOperatingEnvironment}}
\newcommand{\hPhySec}{\dlfont{hasPhysicalSecurity}}
\newcommand{\hPro}{\dlfont{hasProduct}}
\newcommand{\hProGoa}{\dlfont{hasProtectionGoal}}
\newcommand{\hSysAndPlaSec}{\dlfont{hasSystemAndPlatformSecurity}}

\newcommand{\AnaSecRis}{\dlfont{AnalyzedSecurityRisk}}
\newcommand{\Ass}{\dlfont{Asset}}
\newcommand{\Com}{\dlfont{Component}}
\newcommand{\Imp}{\dlfont{Impact}}

\newcommand{\hAnaSecRis}{\dlfont{hasAnalyzedSecurityRisk}}
\newcommand{\hAss}{\dlfont{hasAsset}}
\newcommand{\hCom}{\dlfont{hasComponent}}
\newcommand{\hImp}{\dlfont{hasImpact}}
\newcommand{\hIniSecRisLev}{\dlfont{hasInitialSecurityRiskLevel}}

\newcommand{\ConSecRis}{\dlfont{ControlledSecurityRisk}}

\newcommand{\hResSecRisLev}{\dlfont{hasResidualSecurityRiskLevel}}
\newcommand{\isMitBy}{\dlfont{isMitigatedBy}}

\newcommand{\Assu}{\dlfont{Assumption}}
\newcommand{\Assur}{\dlfont{Assurance}}
\newcommand{\AssurSecDA}{\dlfont{AssuranceSecDA}}
\newcommand{\AssurSecDAI}{\dlfont{AssuranceSecDAI}}
\newcommand{\IntMet}{\dlfont{InterventionMethod}}
\newcommand{\ImpMan}{\IManifest}
\newcommand{\LocOfInt}{\dlfont{LocationOfIntervention}}
\newcommand{\OrgConOfInt}{\dlfont{OrganisationalContextOfIntervention}}
\newcommand{\SecDA}{\dlfont{SecDA}}
\newcommand{\SecDAI}{\dlfont{SecDAI}}
\newcommand{\Str}{\dlfont{Strategy}}
\newcommand{\TesRep}{\dlfont{TestReport}}

\newcommand{\hAssu}{\dlfont{hasAssumption}}
\newcommand{\hAssur}{\dlfont{hasAssurance}}
\newcommand{\hIntMet}{\dlfont{hasInterventionMethod}}
\newcommand{\hImpMan}{\hIManifest}
\newcommand{\hLocOfInt}{\dlfont{hasLocationOfIntervention}}
\newcommand{\hOrgConOfInt}{\dlfont{hasOrganisationalContextOfIntervention}}
\newcommand{\hStr}{\dlfont{hasStrategy}}
\newcommand{\hSubSecDA}{\dlfont{hasSubSecDA}}
\newcommand{\hTesRep}{\dlfont{hasTestReport}}

\newcommand{\LeafSecDA}{\dlfont{LeafSecDA}}
\newcommand{\NotLeafSecDA}{\dlfont{NotLeafSecDA}}
\newcommand{\NotTopSecDA}{\dlfont{NotTopSecDA}}
\newcommand{\TopSecDAAndOneTestReport}{\dlfont{TopSecDAAndOneTestReport}}
\newcommand{\TopSecDA}{\dlfont{TopSecDA}}
\newcommand{\NotTopSecDAAndNoTestReport}{\dlfont{NotTopSecDAAndNoTestReport}}

\newcommand{\LeafSecDAAndOneA}{\dlfont{LeafSecDAAndOneA}}
\newcommand{\NotLeafSecDAAndNoA}{\dlfont{NotLeafSecDAAndNoA}}

\newcommand{\SafImpAss}{\dlfont{SafetyImpactAssessment}}
\newcommand{\SecImpAss}{\dlfont{SecurityImpactAssessment}}

\newcommand{\CriAttPro}{\dlfont{CriticalAP}}
\newcommand{\CriVulLev}{\dlfont{CriticalEL}}
\newcommand{\CriImpLev}{\dlfont{CriticalIL}}
\newcommand{\CriLikFac}{\dlfont{CriticalLF}}

\newcommand{\grt}{\ensuremath{\dlfont{gt}}}
\newcommand\sbullet[1][.5]{\mathbin{\vcenter{\hbox{\scalebox{#1}{$\bullet$}}}}}

\newcommand{\IManifest}{\dlfont{ImplementationManifest}}

\newcommand{\CRL}{\dlfont{CriticalRiskLevel}}

\newcommand{\hIManifest}{\dlfont{hasImplementationManifest}}

\newcommand{\ex}[1]{\exists #1.\top}
\newcommand{\gcialign}[2]{\ensuremath{#1 &\sqsubseteq #2}}

\newcommand{\dom}[1]{\mathit{dom}(#1)}
\newcommand{\ran}[1]{\mathit{ran}(#1)}
\newcommand{\tra}[1]{\mathit{tra}(#1)}

\DeclareFloatingEnvironment[fileext=lol, listname={List of Listings}, name=Listing]{listingi}

\newcommand{\EL}{\ensuremath{\mathord{\mathcal{E}\!\mathcal{L}}}\xspace}
\newcommand{\ELpp}{\ensuremath{\EL^{++}}\xspace}

\newcommand{\Inds}{\mathsf{N_I}}
\newcommand{\Cons}{\mathsf{N_C}}
\newcommand{\Rols}{\mathsf{N_R}}

\newcommand{\dland}{\sqcap}
\newcommand{\dlsub}{\sqsubseteq}

\newcommand{\dlint}{\ensuremath{\mathcal{I}}}

\newcommand{\underint}[1]{\ensuremath{{#1}^\dlint}}

\newcommand{\dldom}{\ensuremath{\underint{\Delta}}}
\newcommand{\dlfunc}{\ensuremath{\underint{\cdot}}}
\newcommand{\Dlint}{\ensuremath{\tuple{\dldom,\dlfunc}}}

\newcommand{\CName}{\ensuremath{\dlfont{A}}}
\newcommand{\RName}{\ensuremath{\dlfont{R}}}
\newcommand{\RNamesub}[1]{\ensuremath{\dlfont{R_{#1}}}}
\newcommand{\IName}{\ensuremath{\dlfont{a}}}
\newcommand{\JName}{\ensuremath{\dlfont{b}}}

\newcommand{\TBox}{\ensuremath{\mathcal{T}}\xspace}

\newcommand{\ABox}{\ensuremath{\mathcal{A}}\xspace}

\newcommand{\inv}[1]{#1^{-}}

\newcommand{\rassert}[3]{\mbox{\ensuremath{#3(#1,#2)}}}
\newcommand{\cassert}[2]{\mbox{\ensuremath{#2(#1)}}}

\newcommand{\ind}[1]{\dlfont{#1}}

\newcommand{\attcount}{\pi}
\newcommand{\vulcount}{\sigma}
\newcommand{\impcount}{\tau}

\newcommand{\dlsev}[1][i]{\ind{s_{#1}}}

\newcommand{\dlatt}[1][i]{\ind{a_{#1}}}
\newcommand{\dlvul}[1][i]{\ind{e_{#1}}}
\newcommand{\dlimp}[1][i]{\ind{i_{#1}}}

\newcommand{\avionto}[3]{\ensuremath{\mathcal{K}^{{a}\textsc{-}\mkern-1mu{v}\textsc{-}\mkern-1mu{i}}_{#1,#2,#3}}}
\newcommand{\aviABox}{\ABox_{\attcount,\vulcount,\impcount}}
\newcommand{\aviTBox}{\TBox_{\grt}}
\newcommand{\aviontot}{\avionto{\attcount}{\vulcount}{\impcount}}

\newcommand{\NPlus}{\mathbb{N}^{+}}
\newcommand{\Graph}{\ABox}
\newcommand{\VGraph}{\ensuremath{\Inds(\Graph)}}

\newcommand{\hasatleast}[1][n]{\mathord{\geq_{#1}}}
\newcommand{\hasatmost}[1][n]{\mathord{\leq_{#1}}}
\newcommand{\hasexactly}[1][1]{\mathord{=_{#1}}}
\newcommand{\hasexactlytop}[1]{\hasexactly[1]{#1}.\top}
\newcommand{\hasatleasttop}[1]{\hasatleast[1]{#1}.\top}
\newcommand{\hasatmosttop}[1]{\hasatmost[1]{#1}.\top}
\newcommand{\hasatmosttopt}[2]{\hasatmost[#1]{#2}.\top}

\newcommand{\geval}[1]{\llbracket{#1}\rrbracket^{\Graph}}

\newcommand{\Constraints}{\mathcal{C}}
\newcommand{\Targets}{\mathcal{B}}
\newcommand{\cramalign}{\addtolength{\jot}{-1ex}}

\newcommand{\mysection}[1]{\section{#1}}
\newcommand{\mysubsection}[1]{\subsection{#1}}

\newif\iflong
 \longtrue

\begin{document}

\pagestyle{headings}
\def\thepage{}
\begin{frontmatter}              

\title{Supporting Cybersecurity Risk Management for Medical Devices via the \secuman Ontology and Shapes}

\markboth{}{}

\author[A]{\fnms{Martin} \snm{Diller}
\thanks{Corresponding Author: Martin Diller, martin.diller@tu-dresden.de}},
\author[B]{\fnms{Anne} \snm{Esslinger}},
\author[A]{\fnms{Piotr} \snm{Gorczyca}},
\author[B]{\fnms{Evi} \snm{Hartig}},
\author[C]{\fnms{Lia} \snm{Kacholdt}},
and
\author[A]{\fnms{Hannes} \snm{Strass}}


\address[A]{Faculty of Computer Science, TU Dresden, Germany}
\address[B]{Else Kröner Fresenius Center for Digital Health, Dresden, Germany}
\address[C]{secunet Security Networks AG}

\begin{abstract} We propose the \secuman~ontology and shapes for representing and analysing cybersecurity risk-management documentation for medical devices. Cybersecurity risks are increasingly relevant for connected medical devices and may have direct consequences for patient safety. Current risk-management files are often maintained as semi-structured natural language text, which makes consistency checking, certification review, and reuse difficult. \secuman~provides a formal OWL-based vocabulary for modelling security-risk context, assessment, control measures, and residual-risk evaluation, and uses SHACL constraints to check structural completeness and conformity with the intended documentation model. The ontology is aligned with VDE Spec 90025 and the related \riskman ontology and shapes, while extending their safety-oriented approach to concepts relevant to cybersecurity risk documentation such as threat scenarios, protection goals, attacker profiles, exposure levels, assets, and secure design arguments. \secuman~is intended to support automated first-pass validation, traceability, and integration of cybersecurity and safety risk-management documentation. 
\end{abstract}

\end{frontmatter}

\mysection{Introduction}

Medical devices are safety-critical systems whose failures may harm patients, users, or the environment. Manufacturers must therefore provide documented justification that device-related risks have been systematically identified, evaluated, and controlled, as required in the EU by the Medical Device Regulation~\cite{EUMDR}. In practice, however, risk-management documentation is still often exchanged as semi-structured natural language. This complicates certification and maintenance: notified bodies must inspect large volumes of text, manufacturers must keep risk files consistent across device versions, and reusable parts of risk analyses and risk controls remain difficult to identify and transfer.

To address these problems for safety risk management, the \riskman~ontology and shapes were proposed in~\cite{GorczycaADHHKKM25}. Building on ISO 14971 and VDE SPEC 90025~\cite{ISO14971,VDESpec24}, \riskman~encodes central concepts and structures used in safety-oriented risk-management documentation for medical devices, while its SHACL shapes express structural and semantic constraints relevant to such documentation. The approach is not intended to determine whether a concrete control measure is technically adequate; rather, it supports automated first-pass validation, improved navigation, and more systematic reuse of risk-management documentation.

For connected medical devices, including wearable monitors, implantable systems, and networked clinical equipment, safety-oriented documentation alone is insufficient. Such devices are now an integral part of modern healthcare: they enable continuous patient monitoring, real-time physiological data acquisition, and remote management, thereby improving diagnosis, treatment, and healthcare efficiency~\cite{KangPCL2018}. However, increasing connectivity also introduces cybersecurity risks that may affect patient safety~\cite{SchwartzRCCCCPZ2018}. Cybersecurity risks in healthcare are therefore not merely theoretical. In Europe, ransomware attacks on hospitals have led to cancelled operations and reduced treatment capacity, as observed in incidents in Spain and the United Kingdom~\cite{apnews2023barcelonaCyberattack,campbellHern2024londonHospitalsCyberattack}, while attacks on national health systems have disrupted access to patient data and critical services~\cite{mashinchi2024healthcare}. In the United States, healthcare data breaches have remained at a sustained high level, with more than 700 large breaches reported annually and over 275 million patient records exposed in 2024 alone~\cite{alder2025healthcareDataBreachReport,alder2026healthcareDataBreachStatistics}. At the same time, the cybersecurity threat landscape continues to intensify, with ransomware accounting for more than half of attacks on the health sector~\cite{enisa2025health,europeanCommissionCybersecurityHealthcare}. These developments show that cybersecurity has become a critical aspect of healthcare systems, with direct implications for patient safety.

Regulatory and guidance documents explicitly reflect this connection. The EU MDR requires software-based systems to be developed in accordance with state-of-the-art principles, including information security principles (MDR Annex I, Sections 3 and 17.2). MDCG 2019-16 likewise treats safety and security risks as interdependent and requires cybersecurity risks and risk controls with potential safety impact to be considered within safety risk-management processes, and vice versa~\cite{EUMDR,mdcg2020cybersecurityMedicalDevices}. Similar expectations are reflected in international guidance and standards, including IMDRF principles on medical device cybersecurity, IEC 81001-5-1, and AAMI TIR57, which emphasize the need to manage cybersecurity risks in conjunction with patient-safety considerations across the device lifecycle~\cite{iec2021healthSoftwareSecurity,imdrf2020medicalDeviceCybersecurity,aami2016medicalDeviceSecurityRiskManagement}.

A separate ontology for cybersecurity risk management is needed because security risks are conceptualized differently from safety risks. Safety risk management typically starts from device-related hazards and analyzes how they may lead to hazardous situations and harms. Cybersecurity risk management must instead account for threat scenarios shaped by the operating environment, product security properties, attack types, protection goals, vulnerabilities, attacker capabilities, affected assets, and impacts. These concepts are often handled by different expert groups and documented using different terminology, evaluation methods, and data structures~\cite{martinez2020safety}. A security ontology must therefore preserve cybersecurity-specific concepts while remaining compatible with safety-oriented documentation, so that security--safety dependencies can be made explicit where needed.

In this work, we provide such an ontology, based on a model for risk-management documentation inspired by VDE SPEC 90025 and \riskman, but specialised to cybersecurity using relevant standards and guidance such as AAMI TIR57, AAMI SW96, NIST SP 800-30, and IEC 81001-5-1~\cite{aami2016medicalDeviceSecurityRiskManagement,aami2023sw96,nist2012sp80030r1,iec2021healthSoftwareSecurity}. Specifically, our contributions are: (i) an ontology for representing central concepts and structural elements of cybersecurity risk-management documentation for medical devices; (ii) SHACL shapes expressing constraints on such documentation; and (iii) a \riskman-compatible modelling approach that supports traceability from threat context through risk assessment to risk control and residual-risk evaluation. \secuman~thereby complements \riskman~by covering the security-risk domain and provides a basis for future integration between cybersecurity and safety risk-management documentation.

\mysection{Related Work}
\label{sec:rel-work}

\paragraph{Risk-management ontologies for medical devices.} Several ontologies have been proposed for medical-device regulation, risk-related concepts, or adjacent compliance tasks, but only few target risk-management documentation itself -- see also the discussion in~\cite{GorczycaADHHKKM25}. Uciteli et al.~\cite{UciteliNTSSFKSS17} defined a Risk Identification Ontology embedded in the General Formal Ontology (GFO)~\cite{Herre10,HerreHBHLM06}, but their focus was on identifying risks in time periods surrounding surgical procedures. Kim et al.~\cite{KimPLL19} integrated concepts from IEC~60601-1, IEC~62304, and ISO~14971 in an ontology for medical-software development processes; however, their model is primarily intended to support process integration and standards compliance, rather than automated reasoning over risk-management files. Schütz et al.~\cite{SchutzFW20} developed an ontology for medical devices in Germany with a broader focus on manufacturers, operators, and legal procedures, yet with the aim of
general semantic interoperability, and based on a legal framework that has since been superseded by the EU MDR.


More recent work has explored semantic technologies for neighboring medical-device regulatory tasks: Chattoraj et al.~\cite{ChattorajWJ24} proposed an OWL/RDF knowledge-graph architecture for representing and querying FDA medical-device regulatory rules, while Zhu et al.~\cite{ZhuZXGLTW26} introduced a methodology for constructing a medical-device risk knowledge graph that combines standards, adverse-event data, and large language models.  These works illustrate a continued interest in machine-processable representations for medical-device regulation and risk-related information, but they do not provide an ontology that supports the generation and validation of medical-device risk-management documentation itself. As already mentioned, this work builds on the \riskman ontology~\cite{GorczycaADHHKKM25}, which formalizes safety-oriented risk-management documentation for medical devices using OWL and SHACL. \secuman~adopts this documentation-centered approach and is intended to integrate with \riskman, for example by mirroring its separation into ``context'', ``assessment'', and ``control'' aspects. However, \secuman~targets the security domain: it models concepts required for medical-device security-risk management, including security-specific mitigation strategies represented as secure-design arguments, in contrast to the safe-design arguments used in \riskman.

\paragraph{Cybersecurity and security-risk ontologies in the medical domain.} 
Ontology-based security modelling covers a wide range of concerns, including security taxonomies, vulnerability and attack representation, threat intelligence, operational technology and industrial Internet of Things security, semantic log analysis, and ontology-supported large-language-model applications, as summarized in surveys on these topics, e.g.~\cite{rosa2017security,adach2022security,hollerer2024survey,jarwar2025modelling,LourenccoAFMV25}.   
As to work closely related to our setting, Sills et al.~\cite{SillsRM20} constructed a cybersecurity knowledge graph for medical-device cyber threat intelligence by integrating manufacturer bulletins, CISA Industrial Control Systems Cyber Emergency Response Team alerts, Wikidata, and FDA AccessGUDID data; their focus, however, is on cyber-threat-intelligence aggregation and embedding-based retrieval rather than on a documentation ontology for security-risk management. Hannou et al.~\cite{HannouALC21} introduced SafecareOnto, a cyber-physical security ontology for healthcare infrastructures that models hospital assets, dependencies, impacts, and protections in order to reason about incident propagation, whereas \secuman~targets medical-device security-risk management files and their validation. Dart and Ahmed~\cite{DartA23} propose CYBER-AIDD, a Unified Modeling Language-based ontology and governance framework for cybersecurity resilience in large healthcare providers, but this is a high-level organizational governance model rather than a formal OWL and SHACL framework for machine-checking manufacturer risk-management documentation. Liu et al.~\cite{LiuSU23} presented an ontology-based framework for classifying forensic evidence from Internet of Medical Things devices; this work is security-adjacent and medical-domain-specific, but it concerns post-incident forensic evidence prioritization rather than prospective security-risk analysis and control documentation.

Work on Internet of Medical Things vulnerability modelling and ontology-assisted automation is closer to our setting, but still differs in its target artefacts. Bughio et al.~\cite{BughioCS24} developed an Internet of Medical Things cybersecurity ontology for remote patient monitoring that represents devices, vulnerabilities, exploits, attacks, services, vendors, and affected assets, and in related work proposed a knowledge-graph-based framework for Internet of Medical Things vulnerability detection~\cite{BughioCS24b}; but they are primarily oriented toward infrastructure, vulnerabilities, and threat intelligence rather than toward documentation and validation.  Ghosh et al.~\cite{GhoshSSMKF25} presented CVE-LLM, an ontology-assisted large language model system for vulnerability evaluation in post-market medical-device cybersecurity, using existing cybersecurity knowledge such as CWE and CAPEC via the SEPSES knowledge graph to support artefacts such as affected-status classifications, justifications, comments, and environmental CVSS vectors; this complements \secuman~by showing the usefulness of ontology-enriched automation, but it does not define a formal ontology and SHACL constraint layer for security-risk management files themselves. Finally, Chander et al.~\cite{ChanderHSHG25} proposed an Intelligent Cybersecurity Ontology Framework for Internet of Medical Things-enabled remote patient monitoring that combines ontology-based modelling of devices, dependencies, vulnerabilities, and threats with graph learning, temporal updates, risk scoring, and adaptive mitigation, thereby addressing operational threat detection and response rather than regulatory or engineering documentation.

Overall, the closest related works are those on medical-device cyber threat intelligence, Internet of Medical Things vulnerability modelling, and ontology-assisted vulnerability evaluation~\cite{SillsRM20,BughioCS24,BughioCS24b,GhoshSSMKF25}; these approaches are complementary to \secuman, since they structure vulnerability, threat, asset, or operational-security knowledge, while \secuman~focuses on a lightweight OWL ontology and SHACL shapes for representing, reasoning over, and validating medical-device security-risk management documentation.

\mysection{Background}

\begin{figure}[t]
    \makebox[\textwidth][c]{%
      \includegraphics[width=1\textwidth]{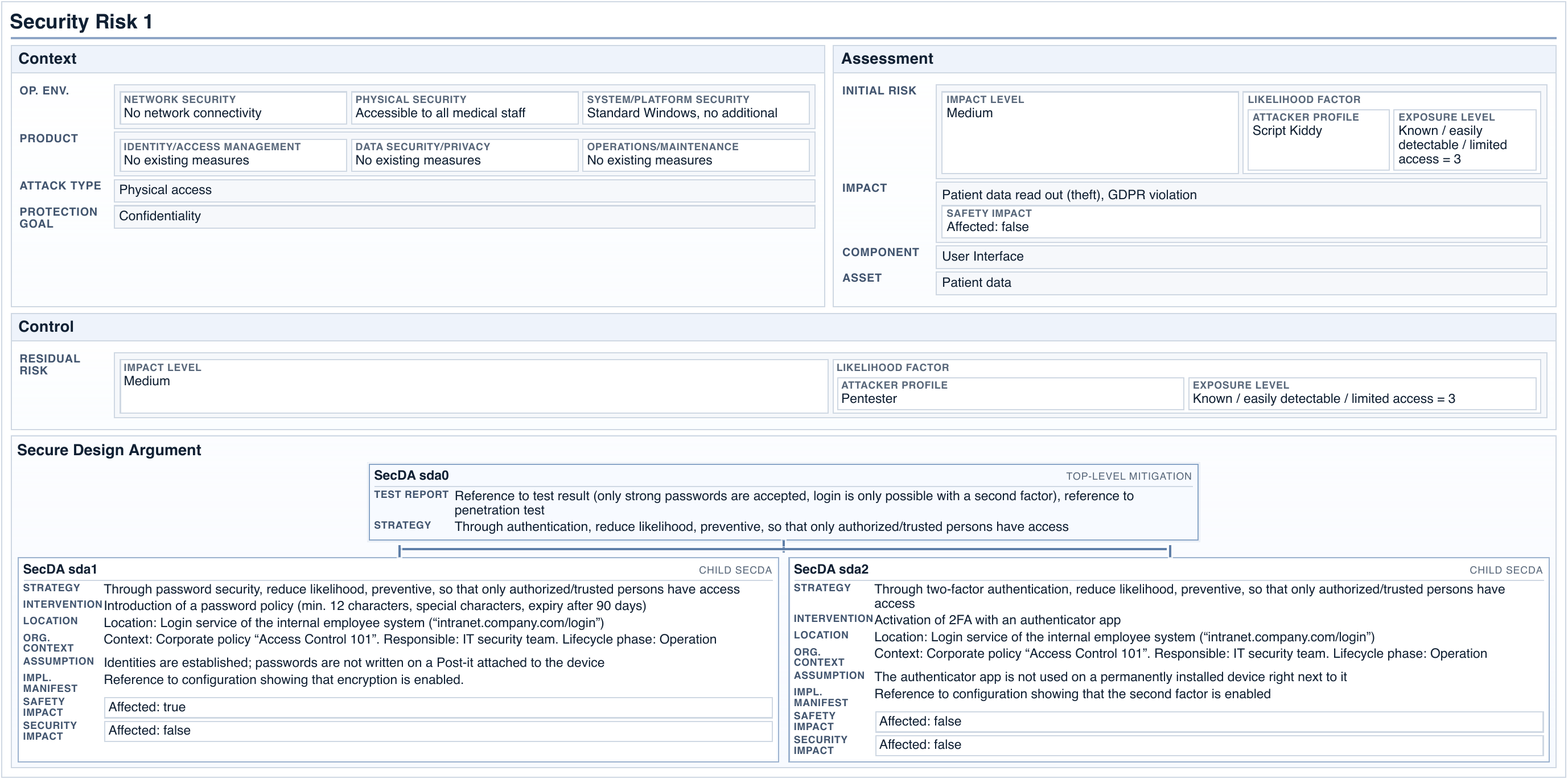}%
    }
    \caption{HTML rendering of a security risk.}
    \label{fig:security-risk-html}
  \end{figure}

  The \secuman ontology follows the \riskman ontology in being based on the recent \vdespeccite, which proposes a structured format for digitalizing risk management files, as well as a machine-readable exchange format using HTML with RDFa~\cite{RDFa}, which annotates selected HTML tags with Resource Description Framework (RDF) triples. However, whereas VDE Spec 90025 and \riskman focus on safety aspects of medical devices, primarily following \isonormcite, \secuman formalises a model for cybersecurity risk management documentation drawing on cybersecurity-relevant standards and guidance, such as AAMI TIR57, AAMI SW96, NIST SP 800-30, and IEC 81001-5-1~\cite{aami2016medicalDeviceSecurityRiskManagement,aami2023sw96,nist2012sp80030r1,iec2021healthSoftwareSecurity}. At the same time, \secuman follows the same philosophy as VDE Spec 90025 and \riskman with respect to the digitalisation and validation of risk management files, and is intended to support integration with safety-focused risk management documentation.

  In contrast to vulnerability-centric approaches, which assess isolated technical weaknesses, the model formalised by \secuman is built around contextualized threat scenarios, where risk emerges from the interaction of the operating environment, device-specific properties, attack types, and attacker capabilities. The model is structured into three sections: Context, Assessment, and Control, mirroring the conceptual structure of safety risk management as formalised in VDE Spec 90025. This separation enables consistent reuse of context attributes across multiple risk entries, thereby reducing the workload involved in creating a security risk management file, and supports traceability from threat description to risk control. The model is designed to support structured documentation of security risks and to enable alignment and systematic integration with safety risk management.
  
  Figure~\ref{fig:security-risk-html} shows the rendering of an HTML file containing a cybersecurity risk structured according to the model formalised by \secuman. The context of the risk describes the conditions under which a threat can occur. It is defined by the combination of information pertaining to the operating environment, product, attack type, and protection goal. These elements represent reusable attributes that are not specific to a single risk, but can be applied across multiple risk scenarios and even across multiple devices employed in similar contexts.
  
  Specifically, the operating environment captures external conditions relevant to cybersecurity. It includes physical security, network security, and system and platform security or hardening. The product describes security-relevant properties of the device itself, including identity and access management, data security and privacy, and operations and maintenance.  The protection goal is one of confidentiality, integrity or availability.  
  
  The assessment aspects describe how the threat affects the system, its operational environment, and relevant stakeholders. In particular, the assessment refers to the context and extends it with the component through which the attack occurs, the affected asset, the impact, and the initial security risk level. The component represents the high-level system element that serves as the entry point of the attack, while the asset represents the physical or digital entity that has value to an individual, an organization, or a government and that is affected by the attack. The impact may include different types of consequences, such as harm to people, financial loss, legal implications, or operational disruption. It also explicitly records whether an impact on safety can occur. The initial security risk level, i.e., the security risk level before any mitigation has been proposed, is composed of an exposure level and an attacker profile, which together form the likelihood factor, as well as an impact level.
  
  The control section represents the risk after the application of control measures. It refers to the assessment aspects and includes a secure design argument and the residual security risk level. The latter has the same structure as the initial security risk level, but represents the risk level after mitigation measures have been applied.
  
  Cybersecurity risk control in the proposed model is represented through secure design arguments (SecDAs). SecDAs are a variant of the safety design arguments proposed in VDE Spec 90025 and formalised in \riskman~-- which in turn are a simplified form of assurance cases as defined in standards such as ISO/IEC/IEEE 15026 and ISO/TS 81001-2-1~\cite{isoIecIeee2019systemsSoftwareAssurance,iso2025healthSoftwareAssuranceCases} -- specialised for mitigation strategies addressing security risks.
  
  SecDAs are structured as trees, with the top-level SecDA representing the claim that a given cybersecurity risk is adequately controlled. Sub-arguments, or sub-SecDAs, contribute to this claim by refining the control strategy into more specific measures. The tree structure enforces a clear separation of abstraction levels: the top-level SecDA defines the overall strategy and includes a test report for effectiveness verification; intermediate SecDAs refine the strategy without implementation detail; and leaf SecDAs define concrete control measures. The strategy defines the overarching approach to risk mitigation, including the control type, i.e., preventive, detective, or corrective, the rationale for the control, and its intended effect on risk parameters, e.g., the likelihood or impact factor. In addition to the strategy, each leaf SecDA contains attributes describing the intervention method, the location of intervention, the organisational context of intervention, assumptions, and an assessment of the impact on safety and security.
  
  Specifically, the intervention method describes the mechanism by which the control is applied. It is the most concrete description of the control and specifies what must be implemented. It may include technical, operational, or administrative measures. The location of intervention specifies where the control is applied within the system, including device internals, connected infrastructure, or, where relevant, the deployment environment. The organisational context of intervention describes the operational setting of the control, including the device lifecycle phase, responsible stakeholders, and relevant policies or practices. The assumptions capture conditions that must hold for the control to be effective, such as the availability of infrastructure or expected user behaviour. Finally, the assessment of the impact on safety and security indicates whether the control measure may itself introduce a safety impact or an additional security impact that requires mitigation.
  
  SecDAs inherit the subcategorisation of safe design arguments from VDE Spec 90025 and \riskman into implementation-specific arguments and assurance arguments. Implementation-specific SecDAs include an implementation manifest, which provides evidence of how the control measure has been realised within the medical device; in particular, leaf SecDAs must always include an implementation manifest. Assurance SecDAs, on the other hand, additionally reference established state-of-the-art practices, such as standards, regulatory guidance, or accepted security frameworks. These references support the justification and traceability of control strategies. In contrast to safety, where standards often define the state of the art more comprehensively, cybersecurity assurance may also rely on evolving best practices and accepted security baselines.

\mysection{Design \& Overview}

\mysubsection{Design}\label{sec:design}
The \secuman{} ontology was developed based on requirements and conceptual design decisions defined by a consortium of experts in medical devices, regulatory affairs, security risk management, ontology engineering, and software development.  
We followed the Linked Open Terms (LOT) methodology~\cite{Poveda-Villalon2022}, which builds on the NeOn methodology~\cite{SurezFigueroa2015TheNM}. The development process comprised three steps:
(1) Requirement specification: We studied relevant norms and standards and examined real medical device security risk management files to establish a shared understanding of the required terms. This was done through regular consortium meetings, workshops, and meetings with manufacturers and security risk experts. The result was a conceptual model of security risk documentation that formed the basis of the \secuman~ontology. (2) Implementation: We encoded the ontology and shapes in OWL and SHACL. During development, we primarily used the HermiT reasoner~\cite{shearer2008hermit} and the OntOlogy Pitfalls Scanner (OOPS!)~\cite{poveda2014oops} to support validity and consistency checks.
(3) Publication and maintenance: The \secuman{} ontology and shapes\footnote{\url{https://github.com/cl-tud/secuman}} as well as the associated validation pipeline\footnote{\url{https://github.com/cl-tud/semeco-q2-validation-pipeline}} are maintained in separate GitHub repositories for version control, issue tracking, and open online access.

\mysubsection{Formal Background}
\label{sec:fml-bkg}

The \secuman{} ontology is encoded in the \EL~profile of \owl~\cite{OWL2Profiles}, while the shapes are specified in \shacl~\cite{SHACL}. For ease of presentation, we describe the ontology here using the description logic \ELpp~\cite{BaaderBL05,BaaderLB08} that underlies the \EL~profile of \owl. For shapes, we use the abstract SHACL syntax introduced by Corman et al.~\cite{CormanRS18,AndreselCORSS20}. Before giving an overview of \secuman{}, we therefore briefly recall the syntax, semantics, and aspects of reasoning relevant to our approach of \ELpp as well as SHACL in the sense of Corman et al.

\paragraph{The Description Logic \texorpdfstring{\ELpp}{EL}} 
\ELpp's concept constructors and their semantics are recalled in the upper part of \Cref{tab:elpp:semantics};
the middle (lower) part shows the constructs allowed in a TBox (ABox).
\begin{table}[h]
    \centering
    \caption{Syntax and semantics of concept, TBox, and ABox expressions of \ELpp.}
    \label{tab:elpp:semantics}
    \setlength{\tabcolsep}{3pt} 
    \noindent
    \bgroup
    \renewcommand{\arraystretch}{0.9}
    \begin{tabular}{l  c  c}
        \toprule
        Name                                    & Syntax                                                   & Semantics                                                                                                                                      \\\midrule
        individual name                         & $\IName\in\Inds$                                         & $\underint{\IName}\in\dldom$                                                                                                                   \\
        concept name                            & $\CName\in\Cons$                                         & $\underint{\CName}\subseteq\dldom$                                                                                                             \\
        role name                               & $\RName\in\Rols$                                         & $\underint{\RName}\subseteq\dldom\times\dldom$                                                                                                 \\
        top                                     & $\top$                                                   & $\dldom$                                                                                                                                       \\
        bottom                                  & $\bot $                                                  & $\emptyset$                                                                                                                                    \\
        nominal                                 & $\set{\IName}$                                           & $\set{\underint{\IName}}$                                                                                                                      \\
        conjunction                             & $C\dland D$                                              & $\underint{C}\cap\underint{D}$                                                                                                                 \\
        existential restriction                 & $\exists\RName.C$                                        & \hspace*{-1em}$\set{x\in\dldom\guard \exists y\in\dldom\colon\tuple{x,y}\in\underint{\RName} \mathbin{\,\&\,} y\in\underint{C}}$\hspace*{-1ex} \\[1pt]\midrule
        \rule{-3pt}{8pt}
        domain restriction                       & $\dom{\RName}\dlsub\CName$                               & $\underint{\RName}\subseteq \underint{\CName}\times \dldom$                                                                                      \\
        range restriction                       & $\ran{\RName}\dlsub\CName$                               & $\underint{\RName}\subseteq\dldom\times\underint{\CName}$                                                                                      \\
        general concept inclusion\hspace*{-1ex} & $C\dlsub D $                                             & $\underint{C}\subseteq\underint{D}$                                                                                                            \\
        role inclusion axiom                    & $\RNamesub{1}\circ\dotsb\circ\RNamesub{k} \dlsub \RName$ & $\underint{\RNamesub{1}}\circ\dotsb\circ\underint{\RNamesub{k}}\subseteq\underint{\RName}$                                                     \\\midrule
        \rule{-3pt}{8pt}
        concept assertion                       & $\cassert{\IName}{\CName}$                               & $\underint{\IName}\in\underint{\CName}$                                                                                                        \\
        role assertion                          & $\rassert{\IName}{\JName}{\RName}$                       & $\tuple{\underint{\IName},\underint{\JName}}\in\underint{\RName}$                                                                              \\\bottomrule
    \end{tabular}
    \egroup
    \vskip1pt
\end{table}
As usual for description logics, the semantics of \ELpp is defined via \define{interpretations} $\dlint=\Dlint$ with a non-empty \define{domain} $\dldom$ and an \define{interpretation function} $\dlfunc$.
An interpretation $\dlint$ is a \define{model} of an ABox $\ABox$ (TBox $\TBox$) if it satisfies all elements of $\ABox$ ($\TBox$) as per \Cref{tab:elpp:semantics}.
An assertion $\alpha$ is \define{entailed} by $\TBox\cup\ABox$, written $\TBox\cup\ABox\models\alpha$, if every model of $\TBox\cup\ABox$ is a model of $\alpha$.


\paragraph{SHACL} To define \shacl, we conveniently re-use description logic vocabulary, viz., pairwise disjoint sets $\Cons$ of \define{classes}, $\Rols$ of \define{properties}, and $\Inds$ of \define{individuals}.
A finite set $\ABox$ of assertions (an ABox) can then be seen as representing a labelled \define{graph}, with individuals acting as nodes, classes labelling nodes, and properties labelling edges.
The syntax of \define{shape expressions} $\phi$ and \define{path expressions} $E$ is shown in \Cref{fig:shacl-syntax}~(top). 
For the semantics, a given graph (ABox) \ABox with nodes (individuals) $\VGraph$ defines an evaluation function \mbox{$\geval{\cdot}$} that assigns to each path expression $E$ a binary relation \mbox{$\geval{E}\subseteq\VGraph\times\VGraph$}, and to each shape expression $\phi$ a set \mbox{$\geval{\phi}\subseteq\VGraph$} 
via induction as shown in \Cref{fig:shacl-syntax} (bottom).

\begin{table}
    \caption{Syntax and semantics of path and shape expressions.}
    \label{fig:shacl-syntax}
    \hrulefill

    The syntax of path expressions $E$ and shape expressions $\phi$ is given by the grammars
    \begin{align*}
        E \ebnfeq \RName \ebnfalt \RName^- \ebnfalt E\cup E \ebnfalt E\sbullet E \ebnfalt E^*
        \text{ and }
        \phi \ebnfeq \top \ebnfalt \CName \ebnfalt \IName \ebnfalt \phi_1\land\phi_2 \ebnfalt \neg\phi\ebnfalt\hasatleast E.\phi\ebnfalt\forall E.\phi\ebnfalt E=E
    \end{align*}%
    where $n\in\NPlus$, $\CName\in\Cons$, $\IName\in\Inds$, and $\RName\in\Rols$ with $\RName^-$ indicating the inverse of $\RName$.%
    \vspace*{4pt}
    \newcommand{\wideelem}[1]{\ensuremath{\makebox[3cm][l]{$#1$}}}
    \newcommand{\skipbelowline}{\raisebox{2pt}[3ex][0pt]{}}
    \bgroup
    \cramalign
    \begin{align*}
        \hline
        \skipbelowline
        \geval{\RName}                    & =\set{\tuple{a,b}\guard \RName(a,b) \in \Graph}
                                          &
        \geval{E_1\sbullet E_2}           & =\geval{E_1}\circ\geval{E_2}                                                                                                                                                             \\
        \geval{\RName^-}                  & =\set{\tuple{b,a}\guard \RName(a,b) \in \Graph}
                                          &
        \hspace*{-1em}\geval{E_1\cup E_2} & =\geval{E_1}\cup\geval{E_2}
                                          &
        \hspace*{-1ex}\geval{E^*}         & = \left(\geval{E}\right)^*                                                                                                                                                               
        \\
        \geval{\top}                      & =\VGraph                                                                                           & \geval{\CName} & =\set{a\guard \CName(a)\in\Graph} & \geval{\IName} & =\set{\IName} \\
        \geval{\phi_1\land\phi_2}         & =\geval{\phi_1}\cap\geval{\phi_2}                                                                  &
        \geval{\neg \phi}                 & =\VGraph\setminus\geval{\phi}                                                                                                                                                            \\
        \geval{\forall E.\phi}            & = \wideelem{\set{a\guard\forall b:\tuple{a,b}\in\geval{E}\text{ implies }b\in\geval{\phi}}}                                                                                              \\
        \geval{\hasatleast[n] E.\phi}     & = \wideelem{\set{a\guard\card{\set{\tuple{a,b}\in\geval{E}\text{ and }b\in\geval{\phi}}} \geq n}}                                                                                        \\
        \geval{E_1=E_2}                   & = \wideelem{\set{a\guard\forall b:\tuple{a,b}\in\geval{E_1}\text{ iff }\tuple{a,b}\in\geval{E_2}}}                                                                                       \\[2pt]
        \hline
    \end{align*}%
    \vspace*{-4ex}
    \egroup
\end{table}

A \define{shape constraint} is an expression of the form $\CName\leftarrow\phi$, with $\CName\in\Cons$ and $\phi$ a shape expression.
A \define{shape schema} is a pair $\tuple{\Constraints,\Targets}$ where $\Constraints$ is a set of shape constraints and $\Targets$ is a set of \define{target} concept assertions.
Intuitively, a target $\cassert{\IName}{\CName}$ expresses the requirement that $\IName$ be labelled by $\CName$.
Formally, an ABox $\ABox$ is a \define{model} for a set $\Constraints$ of constraints if $\geval{\phi}\subseteq\geval{\CName}$ for all $\CName\gets\phi\in\Constraints$.
An ABox $\ABox$ is \define{validated} against a schema $\tuple{\Constraints,\Targets}$ if there exists a set $\Targets'$ of concept assertions such that
(1) $\Targets\subseteq\Targets'$,
(2) $\Inds(\Targets')\subseteq\Inds(\ABox)$, and
(3) $\ABox\cup\Targets'$ is a model for $\Constraints$.  For brevity, in the following we also use the syntactic abbreviations $\hasatmost E.\phi$ for $\neg(\hasatleast[n+1] E.\phi)$ and $\hasexactly[1] E.\phi$ for $\hasatleast[1] E.\phi\land\hasatmost[1] E.\phi$. 
%
%

\paragraph{Reasoning} For reasoning with respect to the \secuman{} ontology and shapes, we adopt a materialisation-based approach. First, all relevant logical consequences of a given ABox together with the \ELpp ontology are computed; more precisely, we derive all entailed concept and role assertions involving concept, role, and individual names occurring in the input ABox.  
Second, the resulting completed ABox is validated against the SHACL constraints.

The feasibility of this approach in our setting is ensured by three observations. First, the role inclusion axioms of \secuman{} satisfy the syntactic restriction imposed by Baader et al.~\cite[Section~3]{BaaderLB08}, and hence entailment of assertions can be decided in polynomial time. Second, \secuman{} uses existential quantification only on the left-hand side of general concept inclusion axioms. As a consequence, materialisation does not introduce fresh individuals, and thus remains finite. Finally, under the formal semantics of Corman et al., \shacl validation is NP-complete in general, whereas certain positive fragments are PTIME-complete~\cite{CormanRS18}.

\mysubsection{The \secuman Ontology \& Shapes}
\label{sec:secuman}

\begin{figure}[t]
    \makebox[\textwidth][c]{%
      \includegraphics[width=\textwidth]{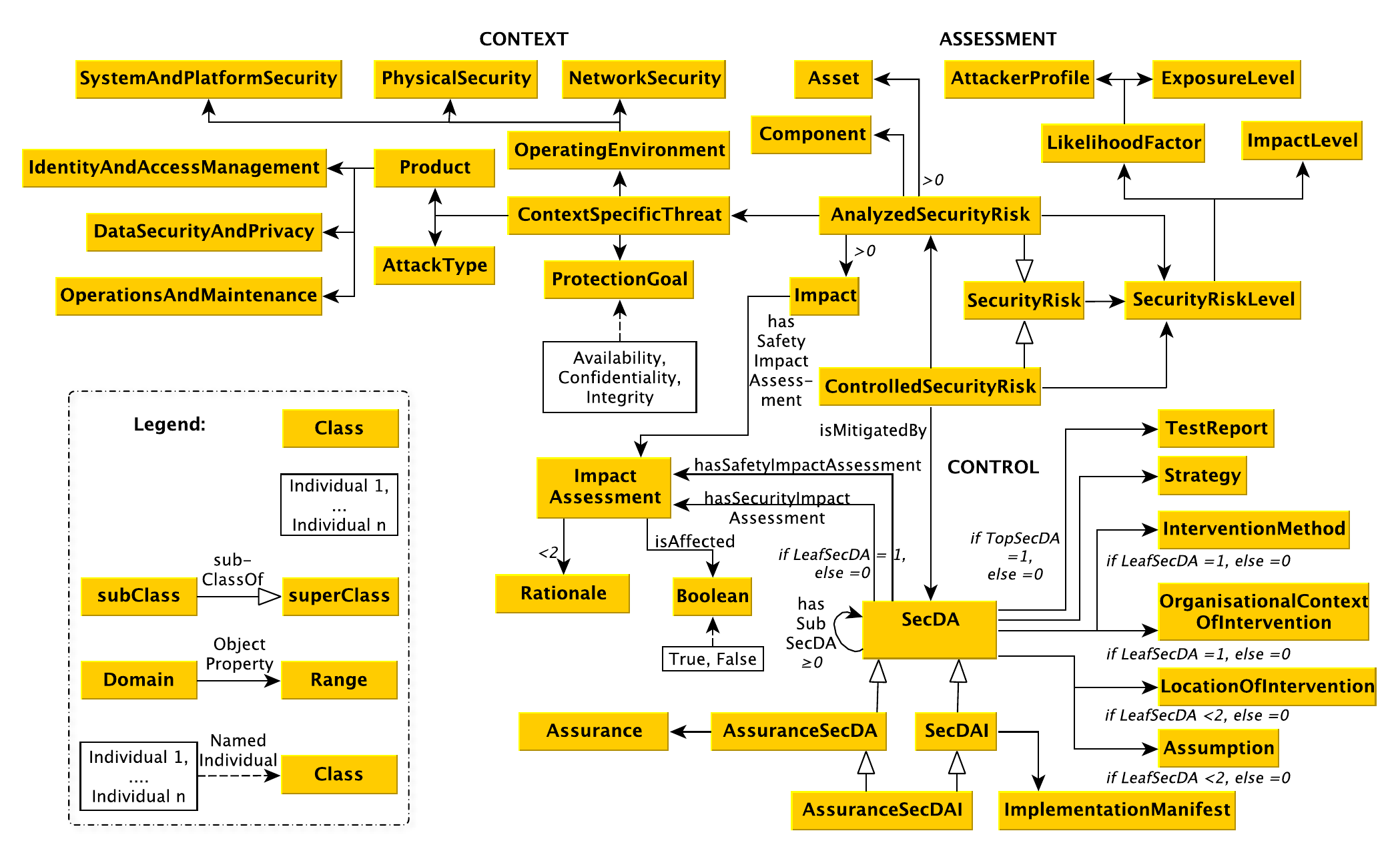}%
    }
    \caption{Schema diagram of the \secuman~classes and properties divided into context, assessment, and control aspects.  Any arrow representing a property without an explicit name means that the property has the name hasX where X is the class in the range, e.g. \hAss~for the property with domain \AnaSecRis~and range \Ass.  Property arrows are also labelled with information pertaining to the Shape constraints, e.g. an \AnaSecRis~can have 1 or more assets or a \SecDA~only has a \TesRep~if it is a top \SecDA~(i.e. is not the child of any other \SecDA), and then it must have exactly one \TesRep.  If there are no such labels, then each instance in the domain must be related to exactly one instance in the range.  Moreover, any two classes without subclass relationship are disjoint.}
    \label{fig:secuman}
  \end{figure}

  \paragraph{Common Aspects} Figure~\ref{fig:general-gci-axioms} presents the GCIs for classes shared across the assessment and control aspects of security risk management documentation. In particular, the central class \SecRis\ is characterized compositionally: every individual with an attacker profile and an exposure level is a \LikFac, every individual with an impact level and a likelihood factor is a \SecRisLev, and every individual with a security risk level is a \SecRis. Finally, there is the GCI that introduces the auxiliary class \ImpAss, which captures an assessment of whether a security risk or a control measure may induce further risks, in particular in the safety or security dimension. Concretely, it states that every individual with an \iAff\ value is an \ImpAss.

    \begin{figure}[H]
    \smaller
    \cramalign
    \begin{align}
    \ex{\iAff}  & \sqsubseteq \ImpAss \label{gci:impass}\\
    \ex{\hAttPro}\sqcap\ex{\hVulLev}  & \sqsubseteq\LikFac\label{gci:likfac} \\
    \ex{\hSecRisLev} &  \sqsubseteq\SecRis\label{gci:secris} \\
    \ex{\hImpLev}\sqcap\ex{\hLikFac}  & \sqsubseteq\SecRisLev\label{gci:secrislev}
    \end{align}%
    \caption{General concept inclusions for common aspects.}
    \label{fig:general-gci-axioms}
    \end{figure}

\noindent \sloppy The domain and range axioms in Figures~\ref{fig:general-range-domain-axioms-i} and~\ref{fig:general-range-domain-axioms} complement these GCIs by fixing the intended source and target classes of the relevant properties, including \hAttPro, \hVulLev, \hImpLev, \hLikFac, \hSecRisLev, \hRat, \hSafImpAss, \hSecImpAss, and \iAff. In particular, both \hSafImpAss\ and \hSecImpAss\ have domain \ImpAssDom. As shown by further GCIs introduced below, this makes it possible to associate impact assessments both with risks and control measures (security design arguments).  Moreover, the range of \iAff~is a Boolean, while an \ImpAss~individual may also have associated a \Rat, which captures, as its name suggests, the rationale underlying a particular impact assessment.

\begin{figure}[H]
    \smaller
    \cramalign
    \begin{align}
        \dom{\hAttPro} &\sqsubseteq \LikFac \\
        \ran{\hAttPro} &\sqsubseteq \AttPro \\
        \dom{\hImpLev} &\sqsubseteq \SecRisLev \\
        \ran{\hImpLev} &\sqsubseteq \ImpLev 
    \end{align}%
    \caption{Range and domain restrictions for common aspects, part~I.}
    \label{fig:general-range-domain-axioms-i}
\end{figure}

\begin{figure}[H]
    \smaller
    \cramalign
    \begin{align}
        \dom{\hLikFac} &\sqsubseteq \SecRisLev \\
        \ran{\hLikFac} &\sqsubseteq \LikFac \\
        \dom{\hRat} &\sqsubseteq \ImpAss \\
        \ran{\hRat} &\sqsubseteq \Rat \\
        \dom{\hSafImpAss} &\sqsubseteq \ImpAssDom \\
        \ran{\hSafImpAss} &\sqsubseteq \ImpAss \\
        \dom{\hSecImpAss} &\sqsubseteq \ImpAssDom \\
        \ran{\hSecImpAss} &\sqsubseteq \ImpAss \\
        \dom{\hSecRisLev} &\sqsubseteq \SecRis \\
        \ran{\hSecRisLev} &\sqsubseteq \SecRisLev \\
        \dom{\hVulLev} &\sqsubseteq \LikFac \\
        \ran{\hVulLev} &\sqsubseteq \VulLev \\
        \dom{\iAff} &\sqsubseteq \ImpAss \\
        \ran{\iAff} &\sqsubseteq \Boolean
    \end{align}%
    \caption{Range and domain restrictions for common aspects, part~II.}
    \label{fig:general-range-domain-axioms}
\end{figure}

At the constraint level, the SHACL shapes in Figure~\ref{fig:general-shacl} make these dependencies explicit for instance data. In particular, each \LikFac\ must have exactly one \AttPro\ and exactly one \VulLev, each \SecRis\ must have exactly one \SecRisLev, and each \SecRisLev\ must have exactly one \ImpLev\ and exactly one \LikFac. Moreover, each \ImpAss\ must specify exactly one \iAff\ value and may have at most one \Rat.  
    
\begin{figure}[H]
    \smaller
    \bgroup
    \cramalign
    \begin{align}
        \ImpAss & \leftarrow  \hasexactlytop{\iAff} \land\hasatmosttop{\hRat}\label{shape:impass}   \\
        \LikFac & \leftarrow  \hasexactlytop{\hAttPro} \land\hasexactlytop{\hVulLev} \label{shape:likfac}  \\ 
        \SecRis & \leftarrow  \hasexactlytop{\hSecRisLev}\label{shape:secris}   \\
        \SecRisLev & \leftarrow  \hasexactlytop{\hImpLev} \land\hasexactlytop{\hLikFac} \label{shape:secrislev}  
    \end{align}
    \egroup
    \caption{Shape constraints for common aspects.}
    \label{fig:general-shacl}
\end{figure}

\paragraph{Context} The GCIs in Figure~\ref{fig:context-gci-axioms}, together with the domain and range restrictions in Figure~\ref{fig:context-range-domain-axioms} capture the context-specific information used to characterize a security concern more precisely. The central class \ConSpeThr\ brings together four dimensions: the operating environment, the product, the attack type, and the protection goal. The two classes \OpeEnv\ and \Pro\ further structure this information: \OpeEnv\ groups physical, network, and system/platform security aspects, whereas \Pro\ groups identity and access management, data security and privacy, and operations and maintenance. The protection goals are given by the distinguished values \Ava, \Con, and \Int.

\begin{figure}[H]
    \smaller
    \cramalign
    \begin{align}
        \begin{split}\label{gci:conspethr}
            \ex{\hOpeEnv}\sqcap\ex{\hPro} & \\
            \sqcap \ex{\hAttTyp}\sqcap\ex{\hProGoa} & \sqsubseteq \ConSpeThr
        \end{split} \\
        \begin{split}\label{gci:opeenv}
            \ex{\hPhySec}\sqcap\ex{\hNetSec} & \\
            \sqcap \ex{\hSysAndPlaSec} & \sqsubseteq \OpeEnv
        \end{split} \\
        \begin{split}\label{gci:pro}
            \ex{\hIdeAndAccMan} & \\
            \sqcap \ex{\hDatSecAndPri} & \\
            \sqcap \ex{\hOpeAndMai} & \sqsubseteq \Pro
        \end{split} \\
        \begin{split}\label{ca:progoa}
            \cassert{\Ava}{\ProGoa} &\quad \cassert{\Con}{\ProGoa} \\
            \cassert{\Int}{\ProGoa} &
        \end{split}
    \end{align}%
    \caption{General concept inclusion axioms and concept assertions for context aspects.}
    \label{fig:context-gci-axioms}
\end{figure}

\begin{figure}[H]
    \smaller
    \cramalign
    \begin{align}
        \dom{\hAttTyp} &\sqsubseteq \ConSpeThr \\
        \ran{\hAttTyp} &\sqsubseteq \AttTyp \\
        \dom{\hDatSecAndPri} &\sqsubseteq \Pro \\
        \ran{\hDatSecAndPri} &\sqsubseteq \DatSecAndPri \\
        \dom{\hIdeAndAccMan} &\sqsubseteq \Pro \\
        \ran{\hIdeAndAccMan} &\sqsubseteq \IdeAndAccMan \\
        \dom{\hNetSec} &\sqsubseteq \OpeEnv \\
        \ran{\hNetSec} &\sqsubseteq \NetSec \\
        \dom{\hOpeAndMai} &\sqsubseteq \Pro \\
        \ran{\hOpeAndMai} &\sqsubseteq \OpeAndMai \\
        \dom{\hOpeEnv} &\sqsubseteq \ConSpeThr \\
        \ran{\hOpeEnv} &\sqsubseteq \OpeEnv \\
        \dom{\hPhySec} &\sqsubseteq \OpeEnv \\
        \ran{\hPhySec} &\sqsubseteq \PhySec \\
        \dom{\hPro} &\sqsubseteq \ConSpeThr \\
        \ran{\hPro} &\sqsubseteq \Pro \\
        \dom{\hProGoa} &\sqsubseteq \ConSpeThr \\
        \ran{\hProGoa} &\sqsubseteq \ProGoa \\
        \dom{\hSysAndPlaSec} &\sqsubseteq \OpeEnv \\
        \ran{\hSysAndPlaSec} &\sqsubseteq \SysAndPlaSec
    \end{align}%
    \caption{Range and domain restrictions for context aspects.}
    \label{fig:context-range-domain-axioms}
\end{figure}

\noindent The corresponding SHACL shapes in Figure~\ref{fig:context-shacl} require these structures to be present explicitly in the data. Thus, each \ConSpeThr\ must specify exactly one value for each of its four dimensions, while each \OpeEnv\ and each \Pro\ must provide exactly one value for each of their respective subdimensions. In addition, every \ProGoa\ value must belong to \ProGoaSet.

\begin{figure}[H]
    \smaller
    \bgroup
    \cramalign
    \begin{align}
        \begin{split}\label{shape:conspethr}
            \ConSpeThr & \leftarrow  \hasexactlytop{\hOpeEnv} \land\hasexactlytop{\hPro} \\ 
                & \phantom{\leftarrow} \quad\land\hasexactlytop{\hAttTyp} \land \hasexactlytop{\hProGoa}    
        \end{split} \\
        \begin{split}\label{shape:opeenv}
            \OpeEnv & \leftarrow  \hasexactlytop{\hPhySec} \land\hasexactlytop{\hNetSec}  \\
                   & \phantom{\leftarrow}\quad\land \hasexactlytop{\hSysAndPlaSec}
        \end{split} \\
        \begin{split}\label{shape:pro}
            \Pro & \leftarrow  \hasexactlytop{\hIdeAndAccMan} \\ 
                & \phantom{\leftarrow}\quad\land\hasexactlytop{\hDatSecAndPri}   \\
                & \phantom{\leftarrow}\quad\land \hasexactlytop{\hOpeAndMai}
        \end{split} \\
        \ProGoa & \leftarrow  \ProGoaSet\label{shape:progoa} 
    \end{align}
    \egroup
    \caption{Shape constraints for context aspects.}
    \label{fig:context-shacl}
\end{figure}

\paragraph{Assessment} \sloppy The GCIs in Figure~\ref{fig:assess-gci-axioms} capture the essential structure of analyzed security risks. In particular, every individual that is related to a context-specific threat, an impact, an initial security risk level, a component, and at least one asset is characterized as an \AnaSecRis. Moreover, the property \hIniSecRisLev\ is specified as a subproperty of \hSecRisLev, which implies that every \AnaSecRis\ is also a \SecRis\ (see Figure~\ref{fig:general-gci-axioms}), while \Imp\ is placed under \ImpAssDom, thereby linking impacts -- and indirectly also analyzed risks -- to the common impact-assessment pattern introduced above.  

\begin{figure}[H]
    \smaller
    \cramalign
    \begin{align}
        \hIniSecRisLev & \sqsubseteq \hSecRisLev\label{ria:inisecrislev} \\
        \Imp & \sqsubseteq \ImpAssDom\label{gci:impassdom1} \\
        \begin{split}\label{gci:anasecris}
            \ex{\hConSpeThr}\sqcap\ex{\hImp} & \\
            \sqcap \ex{\hIniSecRisLev}\sqcap\ex{\hCom} & \\
            \sqcap \ex{\hAss} & \sqsubseteq \AnaSecRis
        \end{split}
    \end{align}%
    \caption{General concept and role inclusion axioms for assessment aspects.}
    \label{fig:assess-gci-axioms}
\end{figure}

\noindent Complementing this conceptual characterization, the domain and range restrictions in Figure~\ref{fig:assess-range-domain-axioms} determine the intended typing of these relations by assigning \AnaSecRis\ as their domain and \ConSpeThr, \Imp, \SecRisLev, \Com, and \Ass\ as the corresponding ranges.

\begin{figure}[H]
    \smaller
    \cramalign
    \begin{align}
        \dom{\hAss} &\sqsubseteq \AnaSecRis \\
        \ran{\hAss} &\sqsubseteq \Ass \\
        \dom{\hCom} &\sqsubseteq \AnaSecRis \\
        \ran{\hCom} &\sqsubseteq \Com \\
        \dom{\hConSpeThr} &\sqsubseteq \AnaSecRis \\
        \ran{\hConSpeThr} &\sqsubseteq \ConSpeThr \\
        \dom{\hImp} &\sqsubseteq \AnaSecRis \\
        \ran{\hImp} &\sqsubseteq \Imp \\
        \dom{\hIniSecRisLev} &\sqsubseteq \AnaSecRis \\
        \ran{\hIniSecRisLev} &\sqsubseteq \SecRisLev
    \end{align}%
    \caption{Range and domain restrictions for assessment aspects.}
    \label{fig:assess-range-domain-axioms}
\end{figure}

\noindent The ontological axioms are mirrored at the constraint level by the SHACL shape in Figure~\ref{fig:assess-shacl}, which makes the intended dependencies explicit for data: each \AnaSecRis\ must specify exactly one context-specific threat, exactly one initial security risk level, and exactly one component, while requiring at least one impact and at least one asset.

\begin{figure}[H]
    \smaller
    \bgroup
    \cramalign
    \begin{align}
        \begin{split}\label{shape:anasecris}
            \AnaSecRis & \leftarrow  \hasexactlytop{\hConSpeThr} \land\hasatleasttop{\hImp}  \\
                   & \phantom{\leftarrow}\quad\land \hasexactlytop{\hIniSecRisLev} \land\hasexactlytop{\hCom} \\
                   & \phantom{\leftarrow}\quad\land \hasatleasttop{\hAss}
        \end{split}
    \end{align}
    \egroup
    \caption{Shape constraints for assessment aspects.}
    \label{fig:assess-shacl}
\end{figure}

\paragraph{Control} The main GCI in Figure~\ref{fig:control-gci-axioms} defines controlled security risks: every individual that is related to an analyzed security risk, a residual security risk level, and a secure design argument is characterized as a \ConSecRis. In addition, and analogously to \hIniSecRisLev\ for analyzed risks, \hResSecRisLev\ is introduced as a subproperty of \hSecRisLev, from which it follows that controlled security risks are also security risks.  

\begin{figure}[H]
    \smaller
    \cramalign
    \begin{align}
        \gcialign{\hResSecRisLev}{\hSecRisLev} \label{ria:ressecrislev} \\
        \begin{split}\label{gci:consecris}
            \ex{\hAnaSecRis}\sqcap\ex{\hResSecRisLev} & \\
            \sqcap \ex{\isMitBy} & \sqsubseteq \ConSecRis
        \end{split}
    \end{align}%
    \caption{General concept and role inclusion for control aspects.}
    \label{fig:control-gci-axioms}
\end{figure}

\noindent The domain and range restrictions in Figure~\ref{fig:control-range-domain-axioms} again fix the expected typing of the associated relations, with \ConSecRis\ as domain and \AnaSecRis, \SecRisLev, and \SecDA\ as the corresponding ranges.

\begin{figure}[H]
    \smaller
    \cramalign
    \begin{align}
        \dom{\hAnaSecRis} &\sqsubseteq \ConSecRis \\
        \ran{\hAnaSecRis} &\sqsubseteq \AnaSecRis \\
        \dom{\hResSecRisLev} &\sqsubseteq \ConSecRis \\
        \ran{\hResSecRisLev} &\sqsubseteq \SecRisLev \\
        \dom{\isMitBy} &\sqsubseteq \ConSecRis \\
        \ran{\isMitBy} &\sqsubseteq \SecDA
    \end{align}%
    \caption{Range and domain restrictions for control aspects.}
    \label{fig:control-range-domain-axioms}
\end{figure}

\noindent Finally, the SHACL shape in Figure~\ref{fig:control-shacl} formulates the corresponding constraints at the data level by requiring each \ConSecRis\ to specify exactly one analyzed security risk, exactly one residual security risk level, and exactly one secure design argument.

\begin{figure}[H]
    \smaller
    \bgroup
    \cramalign
    \begin{align}
        \begin{split}\label{shape:consecris}
            \ConSecRis & \leftarrow  \hasexactlytop{\hAnaSecRis} \land\hasexactlytop{\hResSecRisLev}  \\
                   & \phantom{\leftarrow}\quad\land \hasexactlytop{\isMitBy} 
        \end{split}
    \end{align}
    \egroup
    \caption{Shape constraints for control aspects.}
    \label{fig:control-shacl}
\end{figure}

\paragraph{Secure Design Arguments} The GCIs in Figure~\ref{fig:secda-gci-axioms} define the central structure of secure design arguments and their refinements. In particular, every individual that is related to a strategy is characterized as a \SecDA, while every \SecDA\ with an implementation manifest is classified as a \SecDAI. The remaining GCIs introduce the assurance-specific specializations \AssurSecDA\ and \AssurSecDAI. In addition, the axiom \(\SecDA \sqsubseteq \ImpAssDom\) places secure design arguments into the common impact-assessment pattern introduced earlier. 

\begin{figure}[H]
    \smaller
    \cramalign
    \begin{align}
       \gcialign{\SecDA\sqcap\ex{\hAssur}}{\AssurSecDA} \label{gci:assursecda} \\
        \gcialign{\SecDAI\sqcap\AssurSecDA}{\AssurSecDAI} \label{gci:assursecdai} \\
        \SecDA & \sqsubseteq \ImpAssDom\label{gci:impassdom2} \\
        \begin{split}\label{gci:secsda}
             \ex{\hStr} & \sqsubseteq \SecDA
        \end{split} \\
        \gcialign{\SecDA\sqcap\ex{\hImpMan}}{\SecDAI} \label{gci:secsdai}
    \end{align}%
    \caption{General concept and role inclusion axioms for secure design argument aspects.}
    \label{fig:secda-gci-axioms}
\end{figure}

\noindent The domain and range restrictions in Figure~\ref{fig:secda-range-domain-axioms} fix the expected typing of the relations involving secure design arguments, with \SecDA\ as domain and, depending on the property, assumptions, assurances, intervention methods, implementation manifests, locations and organisational contexts of intervention, strategies, subordinate secure design arguments, and test reports as corresponding ranges. 

\begin{figure}[H]
    \smaller
    \cramalign
    \begin{align}
        \dom{\hAssu} &\sqsubseteq \SecDA \\
        \ran{\hAssu} &\sqsubseteq \Assu \\
        \dom{\hAssur} &\sqsubseteq \SecDA \\
        \ran{\hAssur} &\sqsubseteq \Assur \\
        \dom{\hIntMet} &\sqsubseteq \SecDA \\
        \ran{\hIntMet} &\sqsubseteq \IntMet \\
        \dom{\hImpMan} &\sqsubseteq \SecDA \\
        \ran{\hImpMan} &\sqsubseteq \ImpMan \\
        \dom{\hLocOfInt} &\sqsubseteq \SecDA \\
        \ran{\hLocOfInt} &\sqsubseteq \LocOfInt \\
        \dom{\hOrgConOfInt} &\sqsubseteq \SecDA \\
        \ran{\hOrgConOfInt} &\sqsubseteq \OrgConOfInt \\
        \dom{\hStr} &\sqsubseteq \SecDA \\
        \ran{\hStr} &\sqsubseteq \Str \\
        \dom{\hSubSecDA} &\sqsubseteq \SecDA \\
        \ran{\hSubSecDA} &\sqsubseteq \SecDA \\
        \dom{\hTesRep} &\sqsubseteq \SecDA \\
        \ran{\hTesRep} &\sqsubseteq \TesRep
    \end{align}%
    \caption{Range and domain restrictions for secure design argument aspects.}
    \label{fig:secda-range-domain-axioms}
\end{figure}

Once more, the SHACL constraints in Figure~\ref{fig:secda-shacl-1}, which include several helper constraints,  
specify the structural constraints of secure design arguments at the data level. They require each \SecDA\ to specify exactly one strategy, while allowing at most one location of intervention, at most one assumption, at most one implementation manifest, and at most one assurance. Moreover, the recursive organization of secure design arguments is constrained explicitly: every \SecDA\ must have at least one implementation-level argument reachable via the \hSubSecDA\ relation, which implies in particular that every leaf secure design argument must have an implementation manifest. In addition, top-level arguments must have exactly one test report, whereas non-top-level arguments must have none, and only leaf arguments may carry impact and intervention related information as well as assumptions. For assurance-specific arguments, the corresponding SHACL shape further requires exactly one assurance and propagates the assurance character recursively to subordinate arguments.  Moreover, the shapes for \SecDAI\ and \AssurSecDAI\ require exactly one implementation manifest.  In this respect, the subdivision of secure design arguments into assurance-specific and implementation-specific arguments, as well as the corresponding constraints, parallels that of safe design arguments from~\cite{GorczycaADHHKKM25}; for convenience, however, these structures are duplicated explicitly in the \secuman{} shapes.

\begin{figure}[H]
    \smaller
    \bgroup
    \cramalign
    \begin{align}
        \TopSecDA
            & \leftarrow \hasatmosttopt{0}{\inv{\hSubSecDA}}
        \\
        \NotTopSecDA
            & \leftarrow \hasatleasttop{\inv{\hSubSecDA}}
        \\
        \LeafSecDA
            & \leftarrow \hasatmosttopt{0}{\hSubSecDA}
        \\
        \NotLeafSecDA
            & \leftarrow \hasatleasttop{\hSubSecDA}
        \\
        \TopSecDAAndOneTestReport
        & \leftarrow \TopSecDA \land \hasexactlytop{\hTesRep}
    \\
    \NotTopSecDAAndNoTestReport
        & \leftarrow \NotTopSecDA \land \hasatmosttopt{0}{\hTesRep} \\
               \LeafSecDAAndOneA 
               & \leftarrow \LeafSecDA \land \hasexactlytop{h}
               \notag\\
               & \phantom{\leftarrow}\text{ for all } (A,h) \text{ in } \notag\\
               & \phantom{\leftarrow}\{
                   (\IntMet,\hIntMet), \notag\\
               & \phantom{\leftarrow}     
                   (\OrgConOfInt, \notag\\
                   & \phantom{\leftarrow}   
                   \hOrgConOfInt),\notag\\
                   & \phantom{\leftarrow}
                   (\SafImpAss,\hSafImpAss),\notag\\
                   & \phantom{\leftarrow}
                   (\SecImpAss,\hSecImpAss)\}\\
           \NotLeafSecDAAndNoA
               & \leftarrow \NotLeafSecDA \land \hasatmosttopt{0}{h}
               \notag\\
               & \phantom{\leftarrow}\text{ for all } (A,h) \text{ in } \notag\\
               & \phantom{\leftarrow}\{
                   (\Assu,\hAssu), \notag\\
                   & \phantom{\leftarrow}
                   (\IntMet,\hIntMet), \notag\\
                   & \phantom{\leftarrow}
                   (\LocOfInt,\hLocOfInt), \notag\\
                   & \phantom{\leftarrow}
                   (\OrgConOfInt, \notag\\
                   & \phantom{\leftarrow}
                   \hOrgConOfInt), \notag\\
                   & \phantom{\leftarrow}
                   (\SafImpAss,\hSafImpAss), \notag\\
                   & \phantom{\leftarrow}
                   (\SecImpAss,\hSecImpAss)\} \\
                   \AssurSecDA
                   & \leftarrow \forall\hSubSecDA.\AssurSecDA \notag\\
                   & \phantom{\leftarrow}\quad\land \hasexactlytop{\hAssur}
                   \label{shape:assursecda}
               \\
               \AssurSecDAI
                   & \leftarrow \hasexactlytop{\hAssur} \notag\\
                   & \phantom{\leftarrow}\quad\land \hasexactlytop{\hImpMan}
                   \label{shape:assursecdai}
               \\
               \SecDA
            & \leftarrow \hasexactlytop{\hStr} \notag\\
            & \phantom{\leftarrow}\quad\land \hasatmosttop{h}
                \text{ for all } h \text{ in } \notag\\
            & \phantom{\leftarrow\quad\land}
                \{\hLocOfInt,\hAssu, \notag\\
            & \phantom{\leftarrow\quad\land}      
                \hImpMan,\hAssur\} \notag\\
            & \phantom{\leftarrow}\quad\land \exists\hSubSecDA^*.\SecDAI \notag\\
            & \phantom{\leftarrow}\quad\land
                (\NotTopSecDA \lor \TopSecDAAndOneTestReport) \notag\\
            & \phantom{\leftarrow}\quad\land
                (\TopSecDA \lor \NotTopSecDAAndNoTestReport) \notag\\
            & \phantom{\leftarrow}\quad\land
                (\NotLeafSecDA \lor \LeafSecDAAndOneA) \notag\\
            & \phantom{\leftarrow}\quad
                \text{ for all } A \text{ in } \notag\\
            & \phantom{\leftarrow\quad}
                \{
                    \IntMet, \notag\\
            & \phantom{\leftarrow\quad}
                    \OrgConOfInt, \notag\\
            & \phantom{\leftarrow\quad}
                    \SafImpAss, \notag\\
            & \phantom{\leftarrow\quad}
                    \SecImpAss
                \} \notag\\
            & \phantom{\leftarrow}\quad\land
                (\LeafSecDA \lor \NotLeafSecDAAndNoA) \notag\\
            & \phantom{\leftarrow}\quad
                \text{ for all } A \text{ in } \notag\\
            & \phantom{\leftarrow\quad}
                \{
                    \Assu, \notag\\
            & \phantom{\leftarrow\quad}
                    \IntMet, \notag\\
            & \phantom{\leftarrow\quad}
                    \LocOfInt, \notag\\
            & \phantom{\leftarrow\quad}
                    \OrgConOfInt, \notag\\
            & \phantom{\leftarrow\quad}
                    \SafImpAss, \notag\\
            & \phantom{\leftarrow\quad}
                    \SecImpAss
                \}
            \label{shape:secda}
            \\        
            \SecDAI
            & \leftarrow \hasexactlytop{\hImpMan}
            \label{shape:secdai}   
    \end{align}
    \egroup
    \caption{Shape constraints for secure design argument aspects.}
    \label{fig:secda-shacl-1}
\end{figure}


\paragraph{Disjointness} Finally, the disjointness axioms in Figure~\ref{fig:secuman-disjointness} operate at two conceptual levels. First, pairwise disjointness is imposed on the ontology's primitive classes, i.e., classes that are not introduced as subclasses of other domain classes and therefore serve as basic modeling categories. Second, further disjointness axioms are stated for specialized classes that refine more general concepts. In particular, \AnaSecRis\ and \ConSecRis, both subsumed by \SecRis, are required to be disjoint, and \Imp\ and \SecDA, both subsumed by \ImpAssDom, are likewise required to be disjoint. Together, these axioms exclude unintended class overlap both among primitive categories and among their specialized refinements.

\begin{figure}[H]
    \smaller
    \cramalign
    \begin{align}
        A_i \sqcap A_j & \sqsubseteq \bot \text{ for all distinct } A_i,A_j \text{ in } \notag\\
        & \phantom{\sqsubseteq} \text{\{}\Assu,\Assur,\Ass, \notag\\
        & \phantom{\sqsubseteq} \AttTyp,\AttPro,\Com, \notag\\
        & \phantom{\sqsubseteq} \ConSpeThr,\DatSecAndPri, \notag\\ 
        & \phantom{\sqsubseteq} \IdeAndAccMan, \ImpAss, \notag\\
        & \phantom{\sqsubseteq} \ImpAssDom,\ImpLev, \notag\\
        & \phantom{\sqsubseteq} \ImpMan,\IntMet,\notag\\
        & \phantom{\sqsubseteq} \LikFac, \LocOfInt, \notag\\
        & \phantom{\sqsubseteq} \NetSec,\OpeEnv, \notag\\
        & \phantom{\sqsubseteq} \OpeAndMai,\notag\\
        & \phantom{\sqsubseteq} \OrgConOfInt,\notag\\
        & \phantom{\sqsubseteq} \PhySec, \Pro,\ProGoa,\Rat, \notag\\
        & \phantom{\sqsubseteq} \SecRis,\SecRisLev,\Str, \notag\\
        & \phantom{\sqsubseteq} \SysAndPlaSec,\TesRep,\notag\\
        & \phantom{\sqsubseteq} \VulLev \text{\}}
        \label{gci:secuman-disjoint-classes}
        \\
        \AnaSecRis \sqcap \ConSecRis & \sqsubseteq \bot
        \label{gci:anasecris-consecris-disjoint}
        \\
        \Imp \sqcap \SecDA & \sqsubseteq \bot
        \label{gci:imp-secda-disjoint}
    \end{align}%
    \caption{Disjointness axioms.}
    \label{fig:secuman-disjointness}
\end{figure}

\mysection{Usage \& Extensibility}


\makeatletter
\pgfdeclareshape{document}{
    \inheritsavedanchors[from=rectangle] 
    \inheritanchorborder[from=rectangle]
    \inheritanchor[from=rectangle]{center}
    \inheritanchor[from=rectangle]{north}
    \inheritanchor[from=rectangle]{south}
    \inheritanchor[from=rectangle]{west}
    \inheritanchor[from=rectangle]{east}
    \backgroundpath{
        \southwest \pgf@xa=\pgf@x \pgf@ya=\pgf@y
        \northeast \pgf@xb=\pgf@x \pgf@yb=\pgf@y
        \pgf@xc=\pgf@xb \advance\pgf@xc by-10pt 
        \pgf@yc=\pgf@yb \advance\pgf@yc by-10pt
        \pgfpathmoveto{\pgfpoint{\pgf@xa}{\pgf@ya}}
        \pgfpathlineto{\pgfpoint{\pgf@xa}{\pgf@yb}}
        \pgfpathlineto{\pgfpoint{\pgf@xc}{\pgf@yb}}
        \pgfpathlineto{\pgfpoint{\pgf@xb}{\pgf@yc}}
        \pgfpathlineto{\pgfpoint{\pgf@xb}{\pgf@ya}}
        \pgfpathclose
        \pgfpathmoveto{\pgfpoint{\pgf@xc}{\pgf@yb}}
        \pgfpathlineto{\pgfpoint{\pgf@xc}{\pgf@yc}}
        \pgfpathlineto{\pgfpoint{\pgf@xb}{\pgf@yc}}
        \pgfpathlineto{\pgfpoint{\pgf@xc}{\pgf@yc}}
    }
}
\makeatother

\begin{figure}[H]
  \centering
  \hspace*{-0.4ex}%
  \begin{tikzpicture}[
      every node/.style={
              draw,
              align=center,
              font=\scriptsize,
          },
      pipeline/.style={
              minimum width=2.75cm,
              fill=blue!15
          },
      opt/.style={
              dashed
          },
      ontoshape/.style={
              text width=1.35cm,
              fill=yellow!20
          },
      io/.style={
              fill=green!10
          }
  ]

  \node[pipeline] (distiller) {RDF distiller};
  \node[pipeline] (reasoner) [right=18pt of distiller]  {\EL reasoner};
  \node[pipeline] (validator) [right=28pt of reasoner]  {SHACL validator};

  \node[shape=document, io] (input) [above=0.3cm of distiller]  {HTML +\\RDFa submission};

  \node[ontoshape] (ronto) [above=0.3cm of reasoner, xshift=-1cm]  {\secuman\\ontology};
  \node[opt, ontoshape] (aonto) [right=7.5pt of ronto]  {Optional\\ontology};

  \node[ontoshape] (rshape) [above=0.3cm  of validator, xshift=-1cm]  {\secuman\\shapes};
  \node[ontoshape, opt] (ashape) [right=7.5pt of rshape]  {Optional\\shapes};

  \node[shape=document, io] (output) [right=10 pt of ashape]  {Valida-\quad\qquad\\tion report};

  \draw[->] (input) -- (distiller);
  \draw[->] (ronto) -- (ronto.south|-reasoner.north);
  \draw[->, dashed] (aonto) -- (aonto.south|-reasoner.north);
  \draw[->] (rshape) -- (rshape.south|-validator.north);
  \draw[->, dashed] (ashape) -- (ashape.south|-validator.north);

  \draw[->] (distiller) -- (reasoner);
  \draw[->] (reasoner) -- (validator);
  \draw[->] (validator.east) -| (output.south);
  \end{tikzpicture}
  \caption{Overview of the \secuman~validation pipeline.}
  \label{fig:validation-pipeline}
\end{figure}
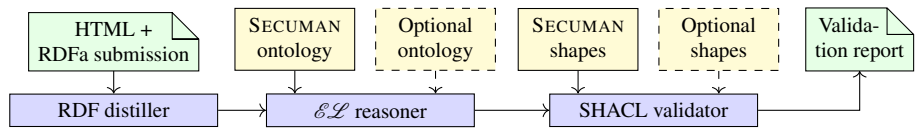

\begin{figure}[t]
  \makebox[\textwidth][c]{%
    \includegraphics[width=1\textwidth]{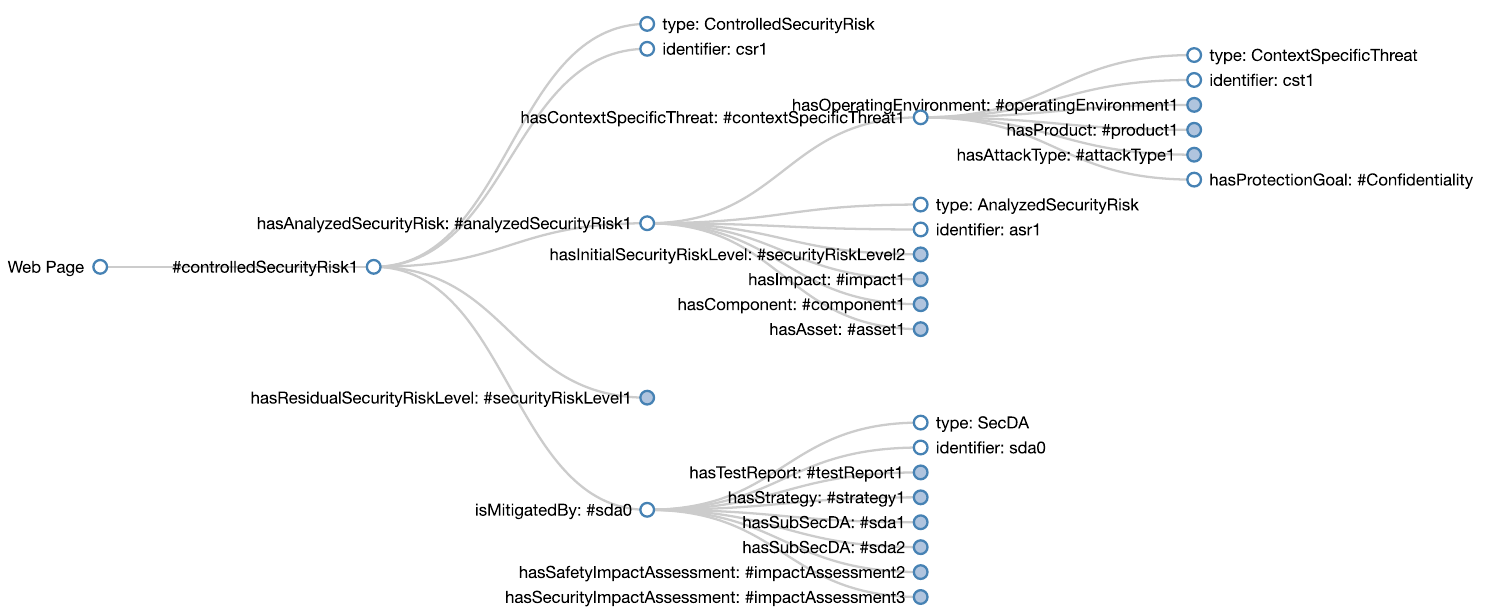}%
  }
  \caption{Partial graphical representation of the RDF for the security risk from Figure~\ref{fig:security-risk-html}.}
  \label{fig:security-risk-rdf}
\end{figure}

Figure~\ref{fig:validation-pipeline} shows the validation pipeline for risk management documentation of medical devices proposed in~\cite{GorczycaADHHKKM25}, which we adapt here for risk management documentation concerning security aspects. As noted above, the approach assumes that risk documentation is submitted as HTML files containing risk management data encoded as RDF triples and expressed using the terminology of the \secuman~ontology. An \emph{RDF distiller} extracts the RDF graph from the HTML by removing the markup and isolating the RDF triples -- see Figure~\ref{fig:security-risk-rdf} for a graphical depiction of the RDF encoding the security risk rendered via HTML in Figure~\ref{fig:security-risk-html}. The RDF triples are then used as input to an OWL (\EL profile)~\emph{reasoner} together with \secuman~and, where needed, additional ontologies. As a result, a materialized knowledge base is obtained and subsequently validated against the \secuman SHACL constraints, and potentially additional constraints, using a \emph{SHACL validator}. The result is then communicated in the form of a human-readable \emph{validation report}.

In our extension of the validation pipeline for \riskman\ (see the URL in Section~\ref{sec:design}), we in particular use the HermiT reasoner~\cite{shearer2008hermit} for materialization, although additional options are provided through a realization wrapper that we also make available online\footnote{\url{https://github.com/cl-tud/realization-wrapper}}. For SHACL validation, we use PySHACL~\cite{SommerA24PySHACL}.

Like \riskman, the \secuman~ontology and shapes are intentionally lightweight so that the validation pipeline can be adapted easily to different modelling needs by incorporating additional ontologies and shapes, as already indicated above. One area in which the ontology will clearly require extension is the modelling of security risk levels, for example attacker profiles, exposure levels, and impact levels, as well as the way in which these are combined to determine whether a risk level is acceptable or unacceptable. A minimalist ``plugin'' for adding such levels to the \secuman~ontology is also provided in our validation pipeline. In this setting, the user may enrich the \secuman~ontology with an ontology \mbox{$\aviontot\mathbin{:\mkern-1mu=}\aviTBox\cup\aviABox$} for $\attcount$ attacker profiles, $\vulcount$ exposure levels, and $\impcount$ impact levels ($\attcount, \vulcount, \impcount \in \mathbb{N}$). This ontology not only introduces individual profiles and levels represented by nominals, but also an ordering $\grt$ over these profiles and levels:

\bgroup
\vspace*{-1ex}
\smaller
\cramalign
\begin{align*}
        \aviABox  =&  \set{\,\cassert{\dlatt}{\AttPro}\guard 1\leq i\leq\attcount} \cup \\
        &  \set{\cassert{\dlvul}{\VulLev}\guard 1\leq i\leq\vulcount} \cup \\ 
        &   \set{\cassert{\dlsev}{\ImpLev}\guard 1\leq i\leq\impcount} \cup         \\
                & \set{\rassert{\dlatt[{i+1}]}{\dlatt}{\grt}\guard 1\leq i<\attcount} \cup \\
                &  
                \set{\rassert{\dlvul[{i+1}]}{\dlvul}{\grt}\guard 1\leq i<\vulcount} \cup \\
                & \set{\rassert{\dlimp[{i+1}]}{\dlimp}{\grt}\guard 1\leq i<\impcount} \\
       \aviTBox =& \set{\tra{\grt}} 
\end{align*}
\egroup%

To encode a concrete security risk matrix -- that is, the way in which risk levels are derived from combinations of attacker profile, exposure level, and impact level -- further axioms would need to be added to $\aviTBox$, since such matrices are typically manufacturer-specific and therefore do not form part of \secuman. As an example, the following axioms first identify critical attacker profiles, exposure levels, and impact levels, and then use these classifications in the last two GCIs to determine critical likelihood factors and, subsequently, critical security levels:

{\smaller
\begin{align*}
&\exists\grt.\{\dlatt[2]\} \sqsubseteq \CriAttPro \\
&\exists\grt.\{\dlvul[3]\} \sqsubseteq \CriVulLev \\
&\exists\grt.\{\dlimp[2]\} \sqsubseteq \CriImpLev \\
&\exists\hAttPro.\CriAttPro\sqcap\exists\hVulLev.\CriVulLev\sqsubseteq \CriLikFac \\
&\exists\hLikFac.\CriLikFac\sqcap\exists\hImpLev.\CriImpLev\sqsubseteq \CRL \\
\end{align*}}

\noindent Checking for controlled security risks with critical residual risk levels can then be achieved by adding the SHACL constraint

{\smaller
\begin{align*}
\ConSecRis \leftarrow \neg(\exists\hResSecRisLev.\CRL)
\end{align*}}

\noindent to the \secuman~shapes. Of course, more complex conditions for deriving and detecting critical security risk levels can likewise be expressed by adding further axioms and shapes.


\mysection{Conclusion}

We presented the \secuman~ontology and accompanying \shacl~shapes for representing and analysing cybersecurity risk-management documentation for medical devices. Analysed risks and their mitigations are represented as an \ELpp~ABox; the \secuman~ontology is used together with an \EL~reasoner to infer implicit knowledge; and \shacl~constraints are used to check whether the resulting data conform to the intended documentation model. The ontology, shapes, and a reference implementation of the complete validation pipeline are freely available.

The \secuman~ontology complements the existing \riskman~ontology by targeting security rather than safety risks of medical devices. It was designed with the ultimate objective of enabling the combined analysis of safety and security risk-management documentation. This objective is reflected not only in the similar structure of the \riskman~and \secuman~ontologies, but also, for instance, in the explicit safety-impact assessments associated with both analysed security risks and SecDAs in the \secuman~ontology. The next step is therefore to develop a bridging ontology for safety and security aspects. Such an ontology should make it possible, for example, to propagate a safety impact identified for a security control measure into the safety-risk-management domain, where the resulting risk can then be analysed and controlled.

Both \riskman~and \secuman~are lightweight ontologies intended to support first-pass validation and, in particular, to facilitate the analysis, navigation, and reuse of medical-device risk-management documentation. More substantive automated analysis will require integration with additional domain ontologies and structured cybersecurity resources; see, e.g.,~\cite{Syed2016UCO,Iannacone2015STUCCO,Kiesling2019SEPSES,Hemberg2020BRON,MITRE_CAPEC,MITRE_CWE,NIST_NVD}, as well as the existing medical-device-specific cybersecurity ontologies and knowledge graphs discussed in Section~\ref{sec:rel-work}. Quality assurance is a particularly significant challenge~\cite{delaVaraJMP19,FosterNGWK21}; possible techniques include dialogue-inspired explication methods from computational argumentation~\cite{DillerGG21,DillerG25}, as well as other approaches for providing justifications in logic-based knowledge-representation formalisms~\cite{DeneckerBS15}.

Finally, the systematic reuse of risk-management documentation remains an important open challenge. We believe that a shared, Semantic Web-based repository of risk-management documentation, including reusable secure and safe design arguments, could substantially accelerate both the creation and the evaluation of such documentation. This potential is especially promising when combined with large language models in retrieval-augmented generation setups~\cite{Lewis2020RAG,Fan2024RAGSurvey}. Realising this vision, however, will require appropriate incentives for sharing risk-management knowledge, as well as infrastructure that supports controlled access and reuse without exposing manufacturer secrets.

\FloatBarrier

\newcommand{\projectname}[1]{\mbox{#1}}
\subsubsection*{Acknowledgements.}
This work was supported by funding from BMFTR (Federal Ministry of Research, Technology and Space) within the project \projectname{SEMECO} (grant no.~03ZU1210BG).

\bibliographystyle{vancouver}
\bibliography{bib/references-imp}

\iflong




\fi

\end{document}